\documentclass[12pt]{article}
\usepackage[reqno]{amsmath}
\usepackage[medium]{titlesec}
\usepackage{amsthm}
\usepackage{bbm}
\usepackage{mathtools}
\usepackage{amssymb}
\usepackage{thmtools}
\usepackage{graphicx}
\usepackage{enumerate}
\usepackage{verbatim}
\usepackage{url}
\usepackage[hidelinks]{hyperref}
\usepackage[dvipsnames]{color}
\usepackage{listings}
\usepackage{authblk}
\usepackage{tabularx}
\usepackage{csquotes}
\usepackage{todonotes}
\usepackage[font=small]{caption}
\usepackage{subcaption}
\usepackage{color}
\usepackage{lscape}
\usepackage{floatrow}
\usepackage{relsize,etoolbox}
\usepackage{pdfpages}
\usepackage{soul}
\usepackage{listings}

\usepackage[backend=bibtex,style=ieee,sorting=none]{biblatex}
\usepackage[margin=2cm]{geometry}

\newcommand{\new}{\ensuremath{\text{new}}}

\begin{document}
\title{\Large\bf Effective sample size: a measure of individual uncertainty in predictions}
\author{\normalsize Doranne Thomassen$^1$, Saskia le Cessie$^{1,2}$, Hans van Houwelingen$^1$, Ewout Steyerberg$^1$}
\affil{\small $^1$Biomedical Data Sciences, Leiden University Medical Center, The Netherlands\\ $^2$Clinical Epidemiology, Leiden University Medical Center, The Netherlands}
\date{\small \today}

\maketitle

\section*{Abstract}
Clinical prediction models are estimated using a sample of limited size from the target population, leading to uncertainty in predictions, even when the model is correctly specified. Generally, not all patient profiles are observed uniformly in model development. As a result, sampling uncertainty varies between individual patients' predictions. We aimed to develop an intuitive measure of individual prediction uncertainty.

The variance of a patient's prediction can be equated to the variance of the sample mean outcome in $n_*$ hypothetical patients with the same predictor values. This hypothetical sample size $n_*$ can be interpreted as the number of similar patients $n_\text{eff}$ that the prediction is effectively based on, given that the model is correct. For generalised linear models, we derived analytical expressions for the effective sample size. In addition, we illustrated the concept in patients with acute myocardial infarction.

In model development, $n_\text{eff}$ can be used to balance accuracy versus uncertainty of predictions. In a validation sample, the distribution of $n_\text{eff}$ indicates which patients were more and less represented in the development data, and whether predictions might be too uncertain for some to be practically meaningful. In a clinical setting, the effective sample size may facilitate communication of uncertainty about predictions. 

We propose the effective sample size as a clinically interpretable measure of uncertainty in individual predictions. Its implications should be explored further for the development, validation and clinical implementation of prediction models.

\section{Introduction}\label{sec:intro}
When using a statistical model to predict a patient's outcome, there is uncertainty in predictions, even when the model is correctly specified. This uncertainty arises from the fact that a prediction model is estimated using a random sample of limited size from the target population. Typically, sampling uncertainty about predictions varies between individual patients, as not all patient profiles are represented uniformly in model development. 

Paradoxically, although a rarer type of patient stands to benefit most from personalised prediction, their predictions are usually more uncertain compared to the average patient. Yet between-patient variation in prediction uncertainty is typically not viewed as a target of optimisation in model development. Neither is it explored as a performance metric in model validation. The TRIPOD guideline for prediction model reporting does not mention any explicit metrics for individual prediction uncertainty \cite{collins2015TRIPOD}.

In the risk communication literature, it has been recommended to present absolute risks in terms of frequencies with a common denominator, preferably with pictograms \cite{trevena2013presenting, fagerlin2011helping}. For example, a patient’s predicted risk can be expressed as a number out of ``100 people like you''\cite{kattan2003_100men}. This presentation fails to capture the uncertainty of the prediction. There is no guideline on the communication of uncertainty around predictions, or whether to communicate uncertainty at all. Statisticians tend to present sampling uncertainty about estimates with a 95\% confidence interval or a standard error. However, communicating sampling uncertainty to patients in the form of confidence intervals has been suggested to have little effect on risk perceptions \cite{Lipkus2001_CIcomm2,han2011_CIComm1} and to increase patient worry \cite{han2011_CIComm1, kattan2011doc}. Another suggestion has been to use verbal qualifiers, such as ``it is most likely that your risk is 30\%" or ``our best estimate is that you have a 40\% risk, though we are not certain about this exact number''. It has also been suggested to communicate numbers only when they are estimated with sufficient certainty \cite{spieg_uncertainty}. 

As an alternative to standard errors and confidence intervals, we aimed to translate the standard error into an ``effective sample size'' being close in spirit to a ``number of patients like you''. This could serve as a more intuitive measure of individual sampling uncertainty in predictions. The effective sample size can then be used as a communication tool in clinical practice and as a performance measure in model validation. In Section \ref{sec:neff}, we define effective sample size in this context and we derive explicit formulas for effective samples sizes in linear and logistic regression. In addition, we illustrate the concept in an example dataset of patients with acute myocardial infarction (Section \ref{sec:examples}). Finally, we discuss the potential applications of effective sample size in prediction modelling and we suggest directions for further investigation (Section \ref{sec:discussion}).

\section{Effective sample size and relative variance}\label{sec:neff}
\subsection{Data setting}
Suppose that for patients $i\in\{1,...,n\}$, we have observed the $n\times 1$ outcome vector $Y$. We have also observed baseline covariates and coded these as desired in the $n\times p$ design matrix $X$. $Y$ and $X$ constitute our \emph{development data}. To the development data, a generalised linear model (GLM) $\mathbb{E}[Y]=g^{-1}(X\beta)$ was fitted. The estimated linear predictor for patient $i$ is denoted as $x_i^\top\hat{\beta}$, where $\hat{\beta}$ is an estimate of $\beta$. After fitting the model to the development sample, it can be used to predict the expected value of $y_\text{new}$ for a new patient, given their covariates $x_\text{new}$. 

\subsection{Effective sample size}
The overall predictive performance of the model can be assessed by comparing predicted and observed values in the development sample (internal validation) or in an external sample of observed patients. A standard validation procedure assesses discrimination and calibration\cite{steyerberg2010assessing}. It does not give any personal measure of uncertainty for a new individual for whom we make a prediction. 

Say there is such a new patient, named Sam. We use the prediction model to predict their expected outcome $y_\text{new}$ based on their covariates $x_\text{new}$. We denote the predicted outcome as $\hat{y}_\text{new}$. To express the individual uncertainty in Sam's prediction, we draw a parallel between the variance of this prediction and the variance of the sample mean $\bar{y}_*$ in a hypothetical independent sample of $n_*$ patients like Sam. Here, being `like Sam' is defined with respect to the model, so having the same design matrix entries as Sam. Suppose that we observe $\bar{y}_*=\hat{y}_\text{new}$ in this hypothetical sample. We define Sam's effective sample size as $n_*$, such that we have as much certainty about Sam's prediction as we would have had, had we directly observed the predicted value in the hypothetical sample of $n_*$ patients like Sam. In other words, given that the model is correct, Sam's prediction is effectively based on $n_*$ similar patients in the development of the model.

The concept of effective sample size can be used as a clinically interpretable measure of individual uncertainty. If the effective sample size for a new patient is small, their covariate profile is exceptional relative to the development data. The development data did not contain much information about patients like them, so the prediction should be interpreted with caution. By contrast, if the effective sample size for a new patient is large, we can have more confidence in their prediction. 

We show in detail how $n_*$ can be derived for linear and logistic regression models and give a general formulation for other generalized linear models. 

\subsection{Linear regression model}
In the linear model, $y_i$ are assumed to be independent samples from a normal distribution with mean $\mu_i=x_i^\top\beta$ and variance $\sigma^2$. The least squares and maximum likelihood estimator for $\beta$ is $\hat{\beta}=(X^\top X)^{-1}X^\top Y,$ whose covariance matrix is $\text{Cov}(\hat{\beta})=\sigma^2 (X^\top X)^{-1}$. Therefore, the variance of a prediction for a new patient, say, Sam, is $\text{Var}(\hat{\mu}_\text{new})=\text{Var}(x^\top_\text{new}\hat{\beta})=\sigma^2x^\top_\text{new}(X^\top X)^{-1}x_\text{new}$. 

Now suppose there was an independent sample of $n_*$ patients who all have the same design matrix entries $x_\text{new}$ and the same prediction $\hat{\mu}_\text{new}$ as Sam. In addition, suppose that we had observed the sample mean $\bar{y}_*$ to be equal to Sam's prediction $\hat{\mu}_\text{new}$. We could use the sample mean $\bar{y}_*$ as an estimator for $\mu_\text{new}$, which would have variance $\sigma^2/n_*$. Solving $\frac{\sigma^2}{n_*}=\text{Var}(\hat{\mu}_\text{new})=\sigma^2x^\top_\text{new}(X^\top X)^{-1}x_\text{new}$ for $n_*$ yields
\begin{align}
n_*&=\frac{1}{x^\top_\text{new}(X^\top X)^{-1}x_\text{new}}.
\end{align}

In the linear model case, the effective sample size is fully determined by the design matrix and does not depend on the outcomes $Y$ through the fitted model. The inverse effective samples sizes for the patients in the development data are equal to the diagonal elements $h_{ii}$ of the hat (projection) matrix, sometimes called leverages. It follows that the harmonic mean of the effective sample sizes in the development data is equal to $n/p$ (where $p$ is the number of parameters in the model).

Usually, leverages are examined as part of model diagnostics. High leverage points have a large potential influence on the fitted model, due to their relatively `extreme' position in the $X$-space. Analogously, the effective sample size of a new prediction expresses how exceptional this new patient's profile is relative to the development data and the fitted model.

\subsection{Logistic regression model}\label{sec:logreg_def}
We now turn to binary outcome variables $y_i$ and assume $y_i\sim\text{Bin}(n,p_i)$, $p_i=\frac{e^{x^\top_i\beta}}{1+e^{x^\top_i\beta}}=g^{-1}(x^\top_i\beta)$, such that $g$ is the logit function. To derive the effective sample size for a new prediction, again for Sam, we require the variance of the prediction $g^{-1}(x^\top_\text{new}\hat{\beta})$. Using the delta method (local linearisation), we have
\begin{align}\label{eq:delta_meth}
\text{Var}(g^{-1}(x^\top_\text{new}\hat{\beta}))&\approx \text{Var}(x^\top_\text{new}\hat{\beta}) \cdot \left(\frac{\delta g^{-1}}{\delta (x^\top_\text{new}\hat{\beta})}\right)^2.
\end{align}
By general properties of the variance, the first factor on the right hand side of \eqref{eq:delta_meth} can be expressed as 
\begin{align}\label{eq:var_linpred}
\text{Var}(x^\top_\text{new}\hat{\beta}) &= x^\top_\text{new}\text{Cov}(\hat{\beta})x_\text{new}.
\end{align}
When $\hat{\beta}$ is the maximum likelihood estimator, $\text{Cov}(\hat{\beta})$ is usually estimated as $(X^\top V X)^{-1}$, where $V$ is a diagonal matrix with elements $v_{ii}=\hat{p}_i(1-\hat{p}_i)$. For the second factor on the right hand side of \eqref{eq:delta_meth}, we have
\begin{align}\label{eq:deriv_g}
\frac{\delta g^{-1}}{\delta (x^\top_\text{new}\hat{\beta})} &= \frac{e^{x^\top_\text{new}\hat{\beta}}}{\left(1+e^{x^\top_\text{new}\hat{\beta}}\right)^2}\\
&=g^{-1}(x^\top_\text{new}\hat{\beta})(1-g^{-1}(x^\top_\text{new}\hat{\beta}))=\hat{p}_\text{new}(1-\hat{p}_\text{new}).
\end{align}
Substituting \eqref{eq:var_linpred} and \eqref{eq:deriv_g} into \eqref{eq:delta_meth} yields
\begin{align}\label{eq:pred_var_logreg}
\text{Var}(g^{-1}(x^\top_\text{new}\hat{\beta}))&\approx x^\top_\text{new}\text{Cov}(\hat{\beta})x_\text{new} \cdot (\hat{p}_\text{new}(1-\hat{p}_\text{new}))^2
\end{align}
for the variance of Sam's prediction.

For a binary variable, the variance is fully determined by the mean. Again, we imagine we have an independent sample of $n_*$ patients with the same design matrix entries $x_\text{new}$ and prediction $\hat{p}_\new$ as Sam. In this sample, we assume to have observed a sample mean of $\bar{y}_*=\hat{p}_\new$. The sample mean $\bar{y}_*$ as an estimator for $\mathbb{E}[y_\new]=p_\new$ would have estimated variance $\frac{\hat{p}_\new(1-\hat{p}_\new)}{n_*}$. Equating this expression to \eqref{eq:pred_var_logreg} and solving for $n_*$ results in an estimate for Sam's effective sample size:
\begin{align}
n_* &\approx \frac{\hat{p}_\text{new}(1-\hat{p}_\text{new})}{x^\top_\text{new}\text{Cov}(\hat{\beta})x_\text{new} \cdot (\hat{p}_\text{new}(1-\hat{p}_\text{new}))^2}\\
&= \left( x^\top_\text{new}\text{Cov}(\hat{\beta})x_\text{new} \cdot (\hat{p}_\text{new}(1-\hat{p}_\text{new}))\right)^{-1}. \label{eq:neff_logreg}
\end{align}
Here, the inverse of the estimated effective sample size for patients in the development data is equal to their corresponding diagonal element of the approximated `hat' matrix (approximated leverage) in logistic regression: $V^{\frac{1}{2}}X\left(X^\top VX\right)^{-1}X^\top V^\frac{1}{2}$\cite{mccullaghGLM}. 

In contrast with the linear model case, the effective sample size depends explicitly on $\hat{\beta}$, and thereby on $Y$, for logistic regression models. It also depends on the predicted risk $\hat{p}$, which will draw the effective sample size towards infinity for patients whose predicted risk is close to 0\% or 100\%. On one hand, it is counterintuitive that extreme extrapolation can lead to exceedingly large effective sample sizes. On the other hand, if a monotonic relation between $x$ and $p$ is assumed, and $p$ is bounded, then it makes sense that certainty about $\hat{p}$ being equal to 0 or 1 increases when $x$ becomes more extreme and the model is correct. 

\subsection{Relative variance}
The inverse of the effective sample size, $n_*^{-1}$, can be interpreted as a \emph{relative variance}, i.e., the prediction variance normalised by the residual or conditional variance. For the linear model, $n_*^{-1}$ is equal to the variance of the prediction $\hat{\mu}_\text{new}$ divided by the residual variance of $y_\text{new}$:
\begin{align}
\text{RelVar} &= n_*^{-1}=x^\top_\text{new}(X^\top X)^{-1}x_\text{new}=\text{Var}(\hat{\mu}_\text{new})/\sigma^2.
\end{align}
For the logistic regression model, we have 
\begin{align}
\text{RelVar} &= n_*^{-1} \approx  x^\top_\text{new}\text{Cov}(\hat{\beta})x_\text{new} \cdot (\hat{p}_\text{new}(1-\hat{p}_\text{new})).
\end{align}
Though less suitable as a communication tool, the relative variance can be used in lieu of effective sample size in model validation. 

\subsection{Other generalised linear models}
For further generalised linear models, an approximation of the effective sample size can be obtained as in Section \ref{sec:logreg_def}. Suppose again that we have a model prediction $\hat{y}_\text{new}=g^{-1}(x^\top_\text{new}\hat{\beta})$ for Sam, based on a fitted GLM, where $g$ is the link function of the GLM. As before, we then assume that we have observed Sam's prediction as the sample mean $\bar{y}_*$ in a hypothetical sample of $n_*$ patients with the same $x$-values as Sam. Given that the model is correct, the sample mean $\bar{y}_*$ has variance $\frac{\text{Var}(Y|x_\text{new}^\top\hat{\beta})}{n_*}$. 

To approximate the prediction variance, we can apply the delta method in \eqref{eq:delta_meth}: $\text{Var}(\hat{y}_\text{new})\approx x^\top_\text{new}(X^\top V X)^{-1}x_\text{new} \cdot \left(\frac{\delta g^{-1}}{\delta (x^\top_\text{new}\hat{\beta})}\right)^2.$ Here, $V$ is a diagonal matrix with diagonal elements $v_{ii}= \left(\frac{\delta g^{-1}}{\delta (x^\top_i\hat{\beta})}\right)^2\left(\text{Var}(Y_i|x_i^\top\hat{\beta})\right)^{-1}$. An expression for Sam's effective sample size follows:
\begin{align}
n_* &\approx \frac{\text{Var}(Y|x_\text{new}^\top\hat{\beta})}{x^\top_\text{new}(X^\top V X)^{-1}x_\text{new} \cdot \left(\frac{\delta g^{-1}}{\delta (x^\top_\text{new}\hat{\beta})}\right)^2}\label{eq:neff_GLM}
\end{align}
For patients in the development data, this expression reduces to the inverse of their approximated leverages $V^{\frac{1}{2}}X\left(X^\top VX\right)^{-1}X^\top V^\frac{1}{2}$. 

\section{Application to GUSTO data}\label{sec:examples}
\subsection{Data description}
We will illustrate the effective sample size in a large data set of patients with acute myocardial infarction (GUSTO-I) \cite{gusto1993}, publicly available from https://hbiostat.org/data/gusto.rda. This data set has been used to illustrate prediction modelling methodology before \cite{SteyerbergBook2019CPMA, steyerberg2004validation, steyerberg2001internal, steyerberg2000prognostic, ennis1998comparison}. The dataset contains data on 40,830 patients, of whom 2851 (7.0\%) died within the first 30 days after myocardial infarction.

\subsection{Illustration with two predictors}
To illustrate the concept of effective sample size, we apply the theory of the previous section to a smaller subset of the GUSTO-I dataset ($n=1214$, comprising study sites from one US region and one non-US region). The variables that we used are DAY30 (a binary indicator of death within 30 days), age, height and shock (a binary indicator of whether the patient was in shock at hospital admission). The variables age and height were centered at their marginal means for the analyses. On average, patients were 61 (SD 11) years old and stood 170 (SD 10) centimeters tall. A minority of 28 (2.3\%) patients presented with shock at admission. 76 patients (6.3\%) died within 30 days after admission. In previous analyses, age and shock were strong predictors of death within 30 days. Height was uncorrelated with age.

For illustrative purposes, we calculated effective sample sizes for two linear regression models. In the first model, the predictors were age and height; in the second model, age and shock were predictors. As effective sample sizes are independent of the outcome in linear regression, the outcome is not relevant here. We also determined effective sample sizes for the following logistic regression models: (1) DAY30 $\sim$ age; (2) DAY30 $\sim$ age + height; (3) DAY30 $\sim$ age + shock.

\begin{figure}
\centering
\begin{subfigure}{.48\linewidth}
\caption{Linear model, $Y\sim$ Age + Height.}
\label{fig:Neff_AGEHEI_OLS}
\includegraphics[width=\linewidth]{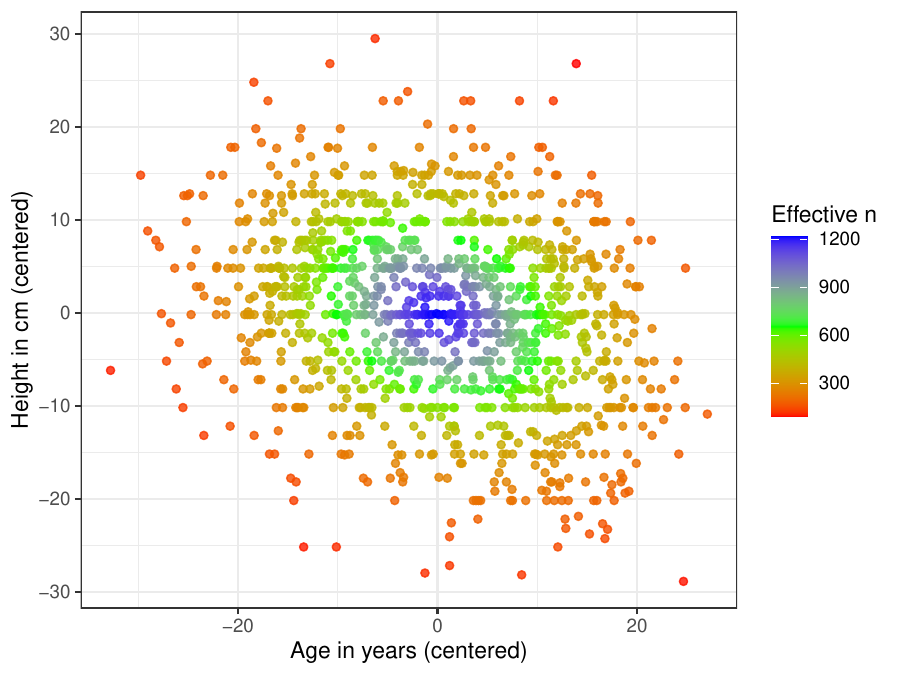}
\end{subfigure}
\begin{subfigure}{.48\linewidth}
\caption{Linear model, $Y\sim$ Age + Shock.}
\label{fig:Neff_AGESHO_OLS}
\includegraphics[width=\linewidth]{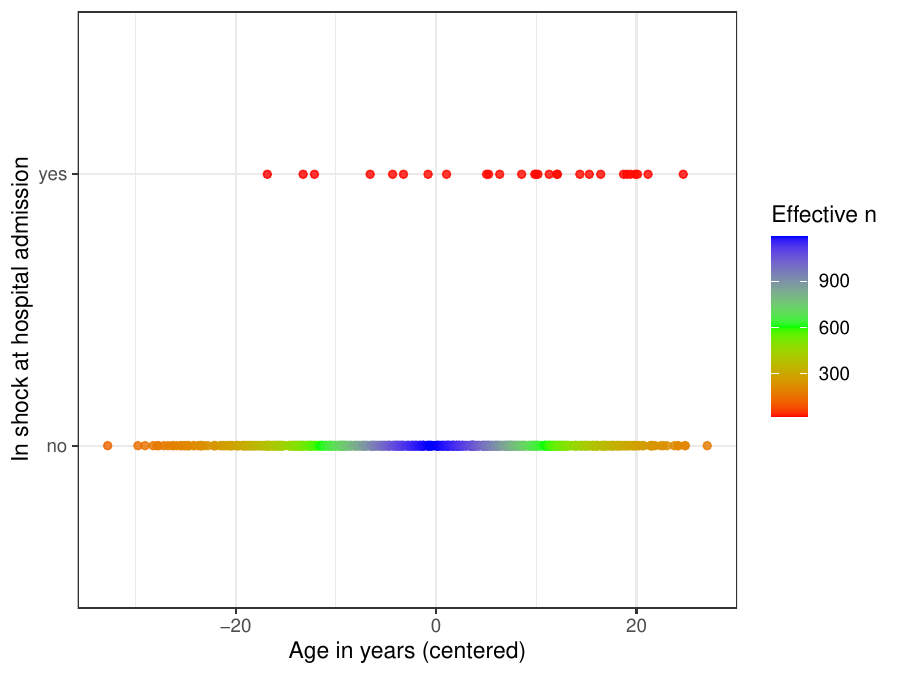}
\end{subfigure}

\begin{subfigure}{.48\linewidth}
\caption{Logistic model, DAY30 $\sim$ Age + Height.}
\label{fig:Neff_AGEHEI_logreg}
\includegraphics[width=\linewidth]{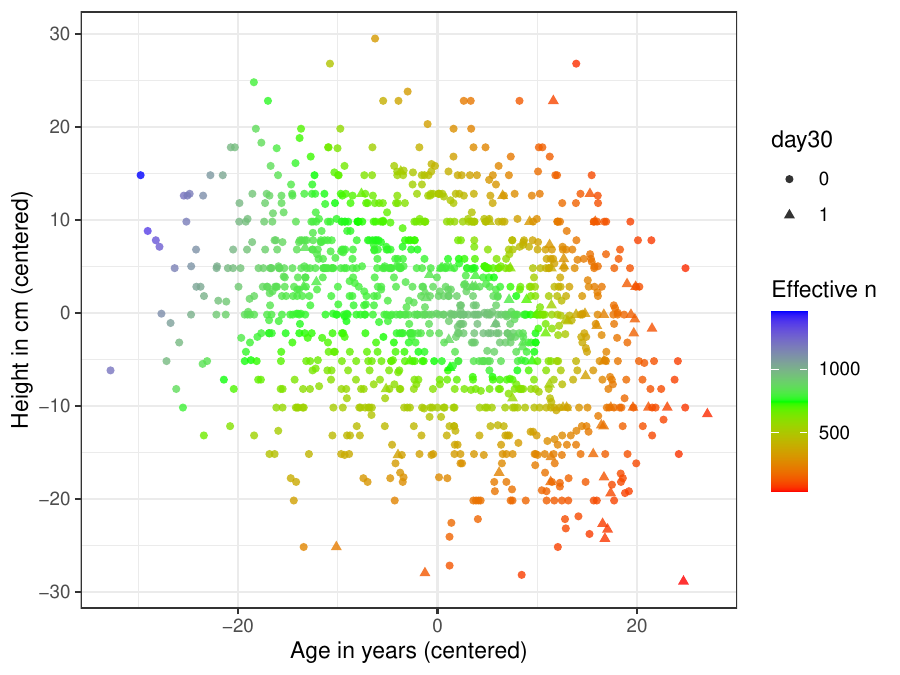}
\end{subfigure}
\begin{subfigure}{.48\linewidth}
\caption{Logistic model, DAY30 $\sim$ Age + Shock.}
\label{fig:Neff_AGESHO_logreg}
\includegraphics[width=\linewidth]{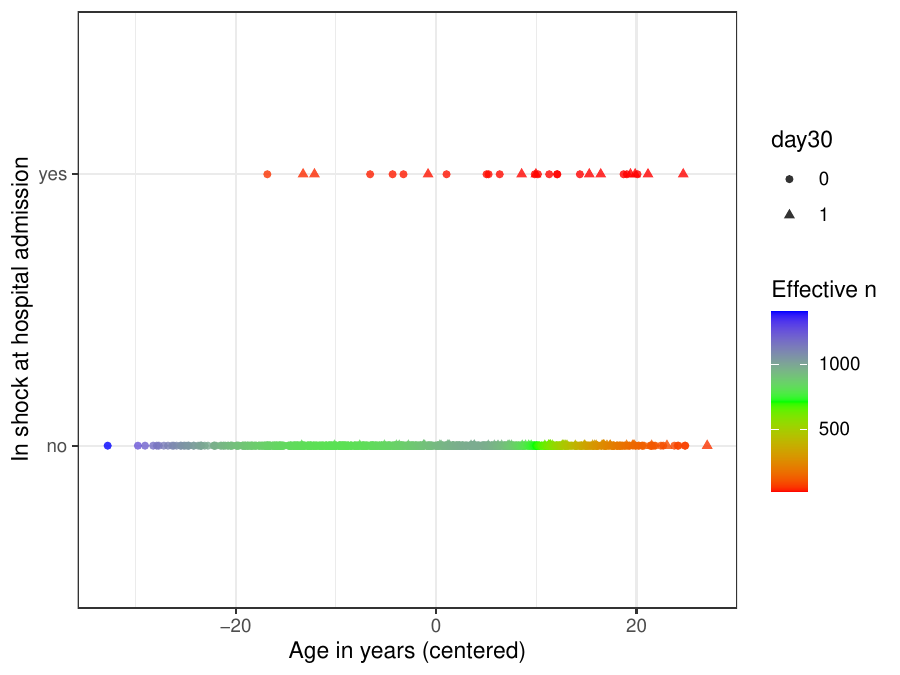}
\end{subfigure}
\caption{Heatmap of effective sample sizes in the GUSTO subsample ($n=1214$) when two-predictor regression models are fitted to predict the outcome of 30-day mortality. Effective sample sizes in linear regression are independent of the outcome, hence the outcome is denoted more generally as $Y$ for these models. In all models, the variables Age and Height were centered at their respective marginal means (61 years and 170 cm). Only a minority of patients in the data presented with shock at admission to hospital.}
\label{fig:Neff_GUSTO1214}
\end{figure}

For the linear regression models, lower effective sample sizes were observed in patients whose covariate values are more `rare' relative to the joint distribution of the covariates (Figures \ref{fig:Neff_AGEHEI_OLS}, \ref{fig:Neff_AGESHO_OLS}). Patients whose covariate values are closer to the average have higher effective sample sizes. Note that in the linear regression case, none of the effective sample sizes exceed the actual total sample size of 1214. 

\begin{figure}
\centering
\begin{subfigure}{.48\linewidth}
\caption{Logistic model, DAY30 $\sim$ Age + Height.}
\label{fig:NP_AGEHEI_logreg}
\includegraphics[width=\linewidth]{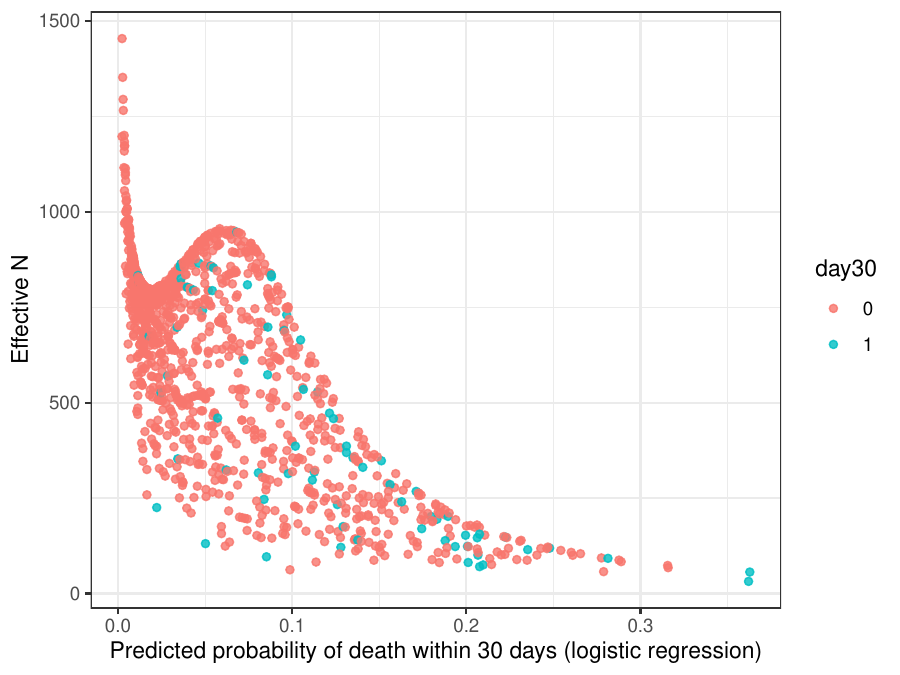}
\end{subfigure}
\hfill
\begin{subfigure}{.48\linewidth}
\caption{Logistic model, DAY30 $\sim$ Age + Shock.}
\label{fig:NP_AGESHO_logreg}
\includegraphics[width=\linewidth]{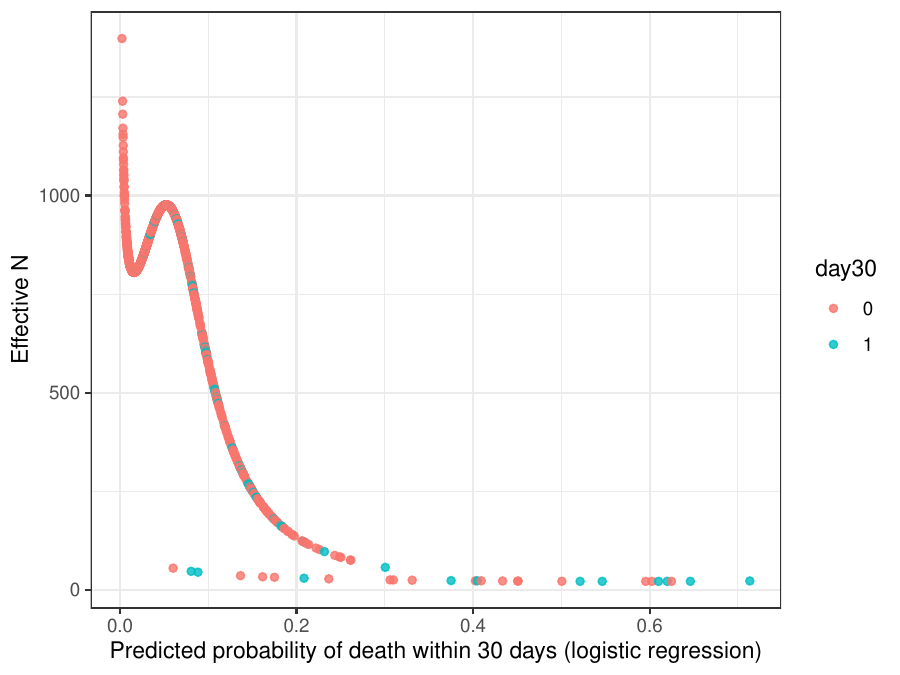}
\end{subfigure}
\caption{Effective sample sizes and the corresponding predicted probabilities of 30-day mortality for patients in the GUSTO subsample ($n=1214$), based on two logistic regression models. In all models, the variables Age and Height were centered at their respective marginal means (61 years and 170 cm). Only a minority of patients in the data presented with shock at admission to hospital.}
\label{fig:NP_GUSTO1214}
\end{figure}

In the logistic models, similar properties appear, though now combined with the impact of the predicted risk being close to zero for some patients (Figures \ref{fig:Neff_AGEHEI_logreg}-\ref{fig:Neff_AGESHO_logreg},\ref{fig:NP_GUSTO1214}). When the predicted risk is close to 0 or 100\%, the effective sample sizes increase, even when the prediction was based on rarer covariate values. This is clearly visible in the upper left corner of Figure \ref{fig:Neff_AGEHEI_logreg}. Younger patients have a lower 30-day mortality risk, which is so close to 0 that their effective sample sizes are actually larger than those of patients whose age was closer to average. Some effective sample sizes even exceed 1214. This illustrates that the effective sample size is not a `real' sample size, but a re-expression of the uncertainty of the prediction. In contrast to linear regression, effective sample sizes for logistic regression may first decrease with extrapolation and then increase again as covariate values become so extreme that predictions approach 0 or 100\%.

The models with shock illustrate effective sample sizes when the model contains a rare binary covariate. From 1214 GUSTO patients, only 28 presented with shock at hospital admission. The effect of shock on death within 30 days is large but uncertain, due to the small number of patients. This is reflected in effective sample sizes for patients with shock, ranging from 25.2 to 28.0 in the linear regression case (Figure \ref{fig:Neff_AGESHO_OLS}). In the logistic regression case, effective sample sizes for these patients ranged from 22.1 to 55.7, as shock is a strong predictor for the outcome and the outcome influences $n_\text{eff}$ here (Figures \ref{fig:Neff_AGESHO_logreg}, \ref{fig:NP_AGESHO_logreg}). Telling a patient with shock that their prediction is based on a study of 1214 patients does not express that there were effectively about 28 patients like them in that study.

\begin{figure}
\includegraphics[scale=0.7]{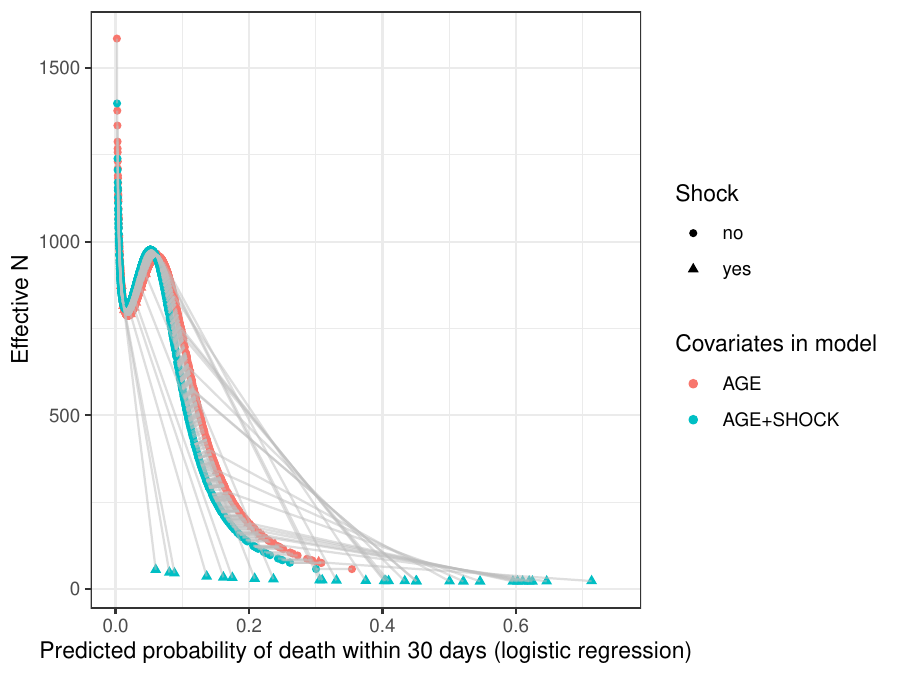}
\caption{Effective sample sizes and the corresponding predicted probabilities for patients in the GUSTO dataset. Two logistic regression models were fitted to the data to obtain predictions: DAY30 $\sim$ Age and DAY30 $\sim$ Age + Shock. Predictions for the same patient are connected by a solid line.}
\label{fig:NP_AGESHO_2_logreg}
\end{figure}

Finally, we visualised a comparison of effective sample sizes in the logistic model with age and shock to a model with age only (Figure \ref{fig:NP_AGESHO_2_logreg}). For the majority of patients without shock, the effective sample size is increased slightly when shock is added to the model, whereas for patients with shock $n_\text{eff}$ decreases dramatically. There is a bias-variance tradeoff here: adding shock to the model may bring predictions closer to the truth for some patients, though at the cost of increased uncertainty. 

\subsection{Development and validation of published prediction models}
To illustrate the use of effective sample sizes in model development and external validation, we used the full GUSTO dataset. We split the data into a development sample of data that was collected within the United States (US, n=23,034) and an external validation sample of data that was collected elsewhere (non-US, n=17,796). 

\begin{figure}
\includegraphics[scale=0.3]{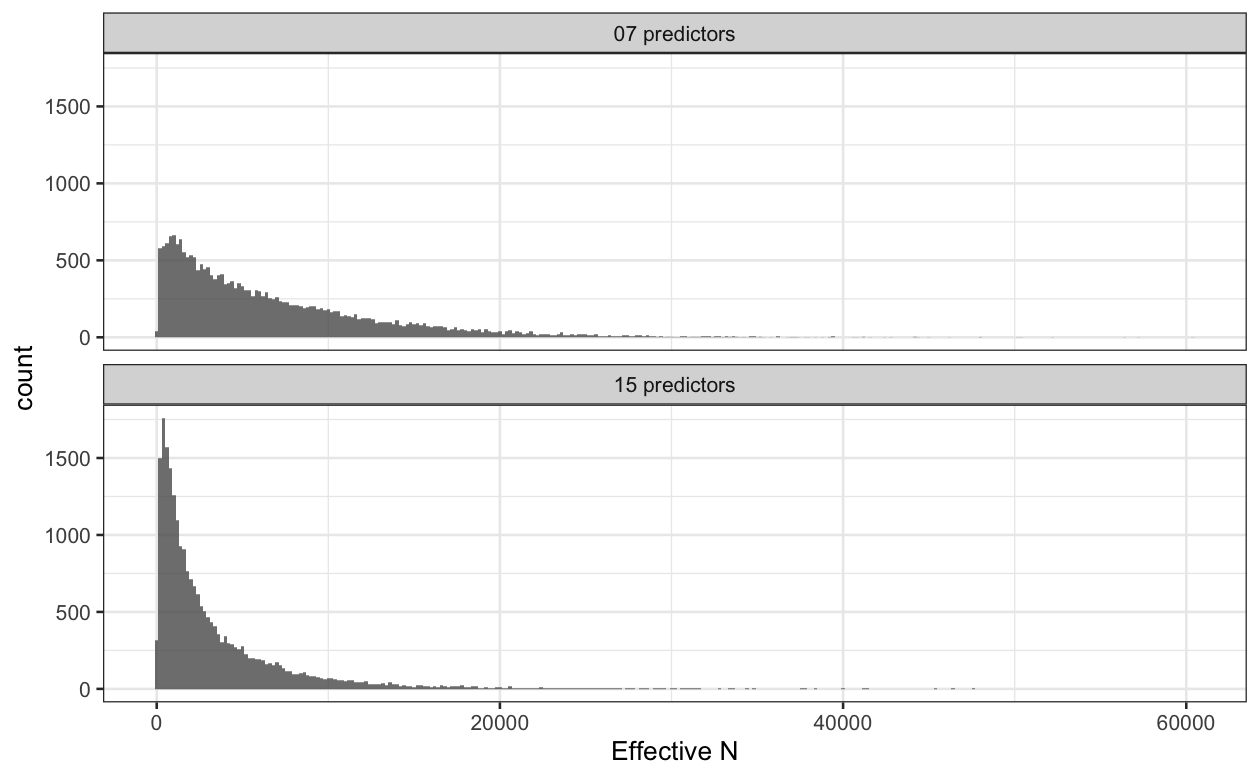}
\caption{Histograms of effective sample sizes in the GUSTO US data (development sample) for a 7- and 15-predictor model for 30-day mortality.}
\label{fig:GUSTO_dev_N_eff_hist}
\end{figure}

Two previously published prediction models were fitted to the US data: one model with 7 predictor variables and 13 model parameters \cite{califf1997}; and a more complex model with 15 predictor variables and 29 model parameters \cite{lee1995}. We then calculated effective sample sizes for each patient in the development (US) data and summarized their distribution (Figure \ref{fig:GUSTO_dev_N_eff_hist}). Lower effective sample sizes are more frequent for the 15 predictor model compared to the 7 predictor model. The total sample size of 23,034 is large, however, hence rarer patient profiles are still quite well-represented in absolute numbers. There were five and sixteen patients with $n_\text{eff}$ below 30 for the 7- and 15-predictor models, respectively.

\begin{figure}
\includegraphics[scale=0.3]{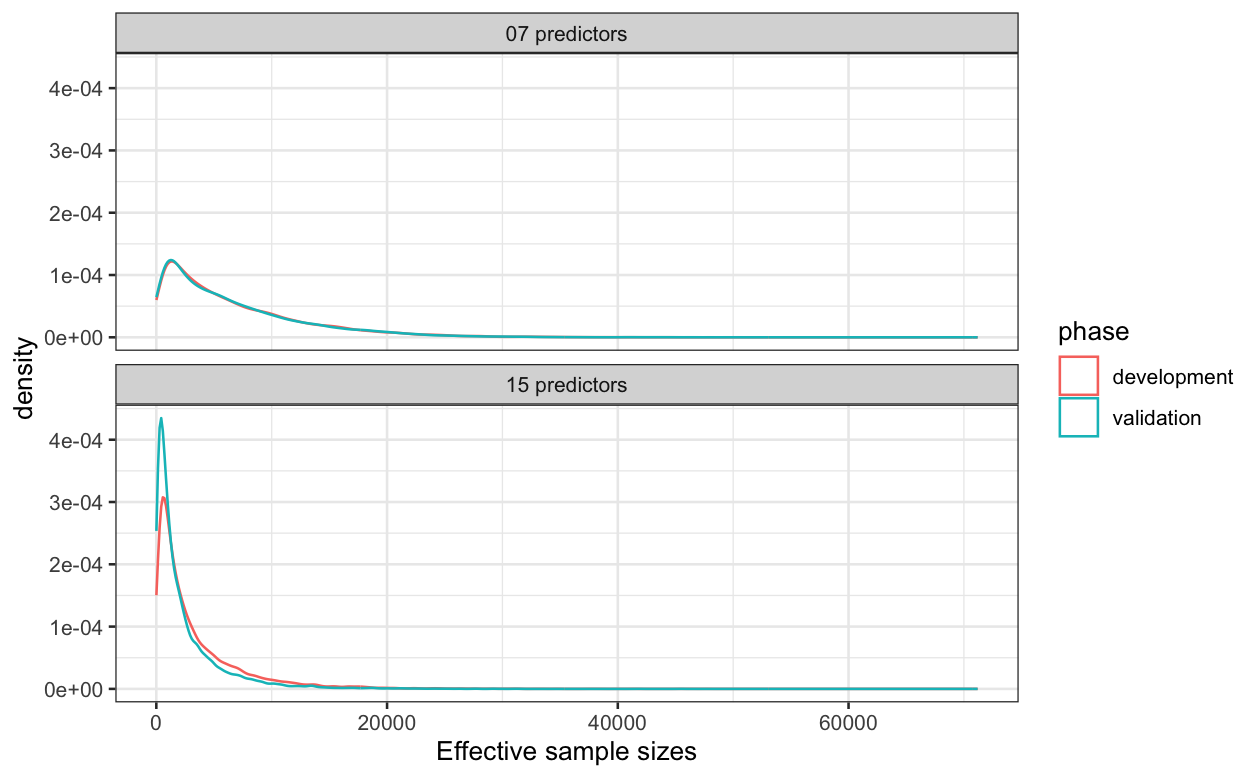}
\caption{Distribution of effective sample sizes in the GUSTO US data (development sample) vs the non-US data (external validation sample) for a 7- and 15-predictor model for 30-day mortality.}
\label{fig:GUSTO_val_Neff_densities}
\end{figure}

We then calculated effective sample sizes for the previously fitted models in the external validation sample (Figure \ref{fig:GUSTO_val_Neff_densities}). For the 7-predictor model, the distribution of effective sample sizes in the validation sample was very similar to that in the development sample. So, with respect to the 7-predictor model the non-US patients overall seem to have a similar prediction uncertainty to the US patients and the occurrence of rare or common covariate patterns seems similar as well. For the 15-predictor model, lower effective sample sizes occur more in the validation sample than they did in the development sample. This indicates that the covariate patterns of patients in the validation sample were somewhat further away from the `average patient' in the development sample.

\section{Discussion}\label{sec:discussion}
We have defined the effective sample size, which can be used as a measure of individual sampling uncertainty in predictions. We derived explicit formulas for the effective sample sizes for linear and generalized linear models. With linear regression models, the effective sample sizes are the inverted leverages and they do not depend on the outcome. In the logistic regression case, effective sample sizes depend on the prediction (and therefore on the outcome) as well as the covariate values, and they increase when the predictions approach 0 or 1. Even with an exceptionally large total development sample size ($>23,000$), we have shown examples in which effective sample sizes dropped below 30 for some patients, indicating large differences in effective representation of patient types in the development of the 7- and 15-predictor models.

We see three directions for application of the concept of effective sample size: in model development and validation; to communicate predicted risks to individual patients; and as a tool to detect patient groups who were underrepresented in model development. In model development, there can be at least two aims regarding the effective sample sizes. The first is that for all patients in the (development/validation) data, their $n_\text{eff}$ is not too low, adding to an overall sample size requirement for prediction model development\cite{riley2020calculating}. This would require specification of a minimal $n_\text{eff}$ for the prediction to be of added value in clinical practice. The minimal effective samples size in model development can be increased by refraining from adding rare binary covariates to a model, like the shock variable in the GUSTO example. There is a bias-variance tradeoff here, however, since shock is a strong predictor for the outcome. Furthermore, $n_\text{eff}$ is defined relative to a specific model and provides a measure of uncertainty conditional on the model. Therefore, it is debatable whether $n_\text{eff}$ can be compared between different models. 

Another objective in model building could be that the effective sample size is more or less the same in all patients, or that their spread remains within certain bounds. Note that shrinkage methods such as Ridge and Lasso regression shrink predictions via the regression coefficients\cite{JamesGareth2021AItS}, which are not individual to each patient. To even out effective sample sizes in a sample of patients, individual shrinkage of the covariates similar to Winsorizing \cite{winsorize} can be applied. This creates a nonlinear relationship between the original covariate and the linear predictor. Further investigation is needed in this direction.

In a model validation setting, the distribution of effective sample sizes in a validation sample summarises individual sampling uncertainty around predictions in a sample of interest. This provides an idea of the variability of prediction uncertainty across patients and patient profiles, as well as a measure of how different (with respect to the model) the validation sample is compared to the development sample.  Therefore, the distribution of effective sample sizes could be an addition to standard performance metrics based on point estimates, such as the c-statistic and calibration summaries \cite{collins2015TRIPOD, steyerberg2010assessing}. 

In addition to the statistical applications of effective sample sizes in prediction, we propose $n_\text{eff}$ as a communication tool in clinical practice. The effective sample size could be used to inform both patients and clinicians about sampling uncertainty in predictions, with statements as ``this number is effectively based on 55 patients like you''. Further empirical research is needed to determine whether $n_\text{eff}$ indeed has beneficial effects on the communication between clinician and patient, and between clinician and model developer. 

Furthermore, effective sample sizes can detect groups of patients to whom we would like to apply a prediction model in clinical practice, but who were effectively underrepresented in the development of the model. We have shown that effective sample sizes can be very small for certain patients, even when the total development sample size was as large as 23,000. Most prediction models are developed with far smaller sample sizes, however. In such a case, effective sample sizes function as a warning sign that the model may not be sufficiently trustworthy to apply in practice for specific groups of patients. When such groups of patients are defined by characteristics historically subject to social biases, detection of their underrepresentation is important in light of recent discussions on algorithmic fairness \cite{Kleinberg2018, Birhane2021, Pessach2023, EU_trustAI}.

In this paper, we focused our derivations on linear and logistic regression models, and provided a generalisation for GLMs. To make the concept of effective sample size more widely applicable, expressions for other types of regression models need to be obtained. In the context of clinical prediction models, the Cox model is a very relevant example. As the variance around a predicted risk from a Cox model is based not only on observed covariates but also on a censoring process, a sample of hypothetical patients ``like'' our new patient is not immediately clearly defined. 

Another direction is to develop a generic numerical algorithm to obtain effective sample sizes when an analytical expression is not available and/or in cases of more complex and flexible models. We suspect that some modelling assumptions are necessary to move from the intuitive definition of $n_\text{eff}$ as `hypothetical sample with the same variance' to a formulation as `prediction variance divided by outcome variance'. If these assumptions can be made, then bootstrap or other sampling-based estimates of the prediction and outcome variances may be used to approximate effective sample sizes. We intend to explore this further, with links to a recently proposed `instability index'\cite{riley2022stability}.

All in all, the concept of effective sample size may prove useful as an intuitive measure of individual prediction uncertainty. Its implications should be explored further for the development, validation and clinical implementation of prediction models.


\subsection*{Financial disclosure}

None reported.

\subsection*{Conflict of interest}

The authors declare no potential conflict of interests.

\subsection*{Data availability statement}
The data that support the results of this paper were loaded from a public repository and are publicly available from https://hbiostat.org/data/gusto.rda.

\section*{Supporting information}
All R\cite{R} code used for the illustrations on the GUSTO data and corresponding output is provided as supporting information.

\newpage
\addcontentsline{toc}{section}{References}
\printbibliography

\end{document}